\documentclass[]{aa}

\usepackage{natbib}
\bibpunct{(}{)}{;}{a}{}{,} % to follow the A&A style
\usepackage{graphicx}
\usepackage{txfonts}
\begin{document}
   \title{Clouds in the atmospheres of extrasolar planets}

   \subtitle{II. Thermal emission spectra of Earth-like planets influenced by low and high-level clouds}

   \author{D. Kitzmann
          \inst{1},
          A.B.C. Patzer
          \inst{1},
          P. von Paris
          \inst{2},
          M. Godolt
          \inst{1},
          \and
          H. Rauer
          \inst{1,2}
          }
          
   \authorrunning{D. Kitzmann et al.}
   \titlerunning{Clouds in the atmospheres of extrasolar planets. II.}

   %\offprints{D. Kitzmann}

   \institute{Zentrum f\"ur Astronomie und Astrophysik, Technische Universit\"at Berlin,
              Hardenbergstr. 36, 10623 Berlin (Germany)\\
              \email{kitzmann@astro.physik.tu-berlin.de}
              \and
              Institut f\"ur Planetenforschung, Deutsches Zentrum f\"ur Luft- und Raumfahrt (DLR),
              Rutherfordstr. 2, 12489 Berlin (Germany)
             }

   \date{Received 02 March 2010 / Accepted 18 May 2011}
% \abstract{}{}{}{}{} 
% 5 {} token are mandatory
 
  \abstract
   {}
   {We study the impact of multi-layered clouds (low-level water and high-level ice clouds) on the thermal emission spectra of Earth-like planets orbiting different types of stars. Clouds have an important influence on such planetary emission spectra due to their wavelength dependent absorption and scattering properties. We also investigate the influence of clouds on the ability to derive information about planetary surface temperatures from low-resolution spectra.}
   {We use a previously developed parametric cloud model based on observations in the Earth's atmosphere, coupled to a one-dimensional radiative-convective steady state climate model. This model is applied here to study the effect of clouds on the thermal emission spectra of Earth-like extrasolar planets in dependence of the type of central star.}
   {The presence of clouds lead in general to a decrease of the planetary IR spectrum associated with the dampening of spectral absorption features such as the $9.6 \ \mathrm{\mu m}$ absorption band of $\mathrm{O_3}$ for example. This dampening is not limited to absorption features originating below the cloud layers but was also found for features forming above the clouds. When only single cloud layers are considered, both cloud types exhibit basically the same effects on the spectrum but the underlying physical processes are clearly different. For model scenarios where multi-layered clouds have been considered with coverages which yield mean Earth surface temperatures, the low-level clouds have only a small influence on the thermal emission spectra. In these cases the major differences are caused by high-level ice clouds. The largest effect was found for a planet orbiting the F-type star, where no absorption features can be distinguished in the low-resolution emission spectrum for high cloud coverages. However, for most central stars, planetary atmospheric absorption bands are present even at high cloud coverages. Clouds also affect the derivation of surface temperatures from low-resolution spectra when fitting black-body radiation curves to the spectral shape of the IR emission spectra. With increasing amount of high-level clouds the derived temperatures increasingly under-estimate the real planetary surface temperatures. Consequently, clouds can alter significantly the measured apparent temperature of a planet as well as the detectability of the characteristic spectral signatures in the infrared. Therefore, planets with observationally derived somewhat lower surface temperatures should not be discarded too quickly from the list of potential habitable planets before further investigations on the presence of clouds have been made.}
   {}


   \maketitle
%
%________________________________________________________________

\section{Introduction}

The climate of Earth-like planets results from the energy balance between absorbed starlight and radiative losses of heat from the planetary surface and atmosphere to space. Clouds reflect starlight back towards space, reducing the stellar energy available for heating the atmosphere (albedo effect), but also reduce radiative losses to space (greenhouse effect). Clouds also have a large effect on the (thermal) emission spectra of planetary atmospheres by either concealing the thermal emission from the surface or dampening the spectral features of so-called biomarker molecules (e.g. $\mathrm{N_2O}$ or $\mathrm{O_3}$). The shape of the IR emission spectra of cloudy atmospheres therefore yields only information about the apparent temperature. However, the apparent temperature can significantly differ from the real surface temperature, which is often used as an important indicator for habitability (possible existence of liquid water).

Directly observed high-resolution mid-IR spectra of the Earth provide a unique opportunity to explore the spectral signatures of habitability in the presence of clouds. According to the studies of such Earth spectra by \citet{Hearty2009} clouds cause a greater spectral variation in the IR observations than the observed differences between day and night. Their findings also indicate that in the case of Earth the main (molecular) spectral features in mid-IR are visible even for cloud covered conditions, but their detectability decreases with lower spectral resolution. \citet{Tinetti2006a,Tinetti2006b} investigated measured and calculated high-resolution spectra of Earth for different cloud conditions to study the influence of clouds. They also found that IR spectra are very sensitive to the coverages and types of clouds. Using measured cloudy Earth spectra \citet{Kaltenegger2007} studied the development of IR emission spectra at different evolutionary stages of Earth assuming the modern Earth cloud cover at all epochs \citep[see also][]{Selsis2008}.

Model calculations of high-resolution IR spectra of Earth-like planets around different stars have been performed by \citet{Segura03,Segura05} for clear sky atmospheres with tuned surface albedos to mimic the climatic effects of clouds. \citet{Selsis2004} summarised the detection and characterisation limitations of terrestrial extrasolar atmospheres without discussing the impact of clouds, concluding that the analysis of the thermal emission at low resolution provides the most comprehensive scientific characterisation of this kind of planetary atmospheres.

In this work we study the impact of multi-layered clouds on low-resolution IR emission spectra of Earth-like planets around different stars. To do this we apply a one-dimensional steady state radiative-convective atmospheric climate model. The model accounts for two different cloud layers, namely low-level water droplet and high-level ice particle clouds, and also for the partial overlap of these two layers. A more detailed model description is given in Sect. \ref{sec_model}. In order to verify the applicability of our modelling approach this coupled cloud-climate model is applied to the modern Earth atmosphere and its spectral appearance as described in Sect. \ref{sec_earth_ref}. Sect. \ref{sec_spectra} presents the resulting thermal IR spectra of cloudy atmospheres of Earth-like planets orbiting different types of central stars and discuss also implications for the detectability of characteristic molecular signatures and the determination of the planetary surface temperatures from spectral shapes.

\section{Details of the model}
\label{sec_model}

We apply a one-dimensional steady state radiative-convective climate model, which accounts directly for the climatic and radiative effects of multi-layered clouds (see \citet{Kitzmann2010a}, henceforth called Paper I, for a detailed model description). In particular the influence of two different cloud layers (low-level water and high-level ice clouds) is included in the model, originally based upon the cloud-free climate model developed by \citet{Kasting1984} and \citet{Pavlov00}. Analytical expressions for the particle size distributions of the two different clouds in Earth-like atmospheres derived from in-situ measurements of the respective cloud types in the Earth atmosphere have been used. Other cloud properties, such as the optical depth and the pressures at the cloud tops, have been taken from satellite measurements. The altitude of each cloud layer is iteratively adjusted to match the corresponding measured pressures and is therefore not simply fixed (see Paper I for a detailed description of the cloud parametrisation).
Other atmospheric feedback effects (e.g. due to different temperature profiles) onto the clouds are not explicitly taken into account. However, the validity range of the approach (freezing limit for water droplet clouds and limiting temperature for liquefying ice clouds) is considered. Further influences on the cloud particles are not taken into account in the parametrised cloud description. Such differences could arise by e.g. considering planets with a different landmass distribution compared to Earth. Therefore, the cloud parametrisations used here are limited to only Earth-like planets. 
\begin{figure}
  \centering
  \resizebox{\hsize}{!}{\includegraphics{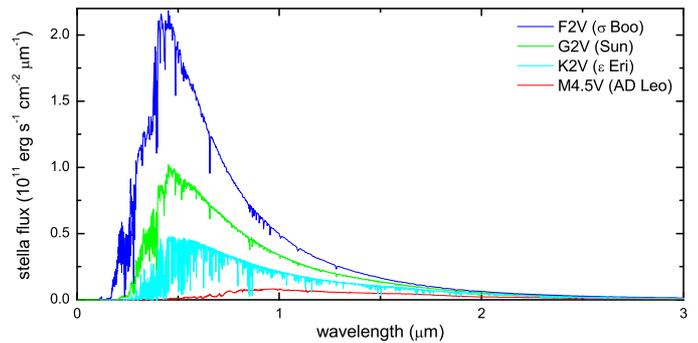}}
  \caption{High-resolution spectra of the different central stars shown for illustrative purposes. The stellar radiation flux is given at the stellar surfaces.}
  \label{stellar_spectra}
\end{figure}
To limit the number of free parameters in the parametrised cloud description the minimum partial overlap of the cloud layers is assumed if the sum of the two cloud coverages exceeds $100\%$. Otherwise non-overlapping cloud layers are considered.
The atmospheric chemical composition, which is assumed for all calculations, is chosen to represent the modern Earth atmosphere and was calculated using a detailed photochemical model \citep[see][]{Grenfell07}. Detailed cloud feedback effects on the chemical composition of the planetary atmospheres are not considered.

The radiative transfer problem is solved by applying a broadband model, which has been optimised for the energy transport in Earth-like planetary atmospheres \citep{Segura03,Mlawer97}. In the IR part 16 spectral bands with variable spectral resolutions ($1<R<40$) are used. The equation of radiative transfer is solved using a hemispheric-mean two stream method including multiple scattering \citep{Toon89}. The gas opacities are described by correlated-k methods \citep[see][and Paper I for details]{Segura03,Mlawer97}, which are an extension of opacity distribution functions \citep{Strom1966,Mihalas1978StellarAtmos}. The correlated-k values derived from the rapid radiative transfer model (RRTM, developed by \citet{Mlawer97}) have been used by \citet{Segura03} for the same planetary scenarios. However, note that the k-distributions used in the RRTM are only tabulated over a limited pressure and temperature range. In particular the temperature range is limited to within $\pm 30 \ \mathrm{K}$ of an Earth mid-latitude summer temperature profile (see \citet{Mlawer97} for details). The application of the k-coefficients beyond this range may lead to inaccuracies, especially for situations strongly deviating from (mean) Earth conditions (see also \citet{Segura03,Segura05} and \citet{vonParis2008}). 

The frequency dependent optical depth effects of the clouds have been incorporated into the two-stream radiative transfer schemes \citep{Toon89} of the atmosphere model.
To account for different amounts of coverages of multi-layered clouds and their partial overlap in our model, we use a flux-averaging procedure, whereby the radiative transfer is solved separately for every distinct cloud configuration. The mean radiative flux which enters into the atmospheric model calculation is obtained by averaging these fluxes weighted with the corresponding cloud cover values.

In our calculations of Earth-like planetary atmospheres four different kinds of central stars are considered: F2V, G2V, K2V, and M4.5V-type stars \citep[see also][]{Segura03,Segura05}. The corresponding stellar spectra are shown in Fig. \ref{stellar_spectra}. The incident stellar fluxes are scaled by varying the orbital distances, such that the energy integrated over each incident stellar spectrum equals the solar constant at the top of the atmospheres of the corresponding Earth-like planets. All four stellar spectra (Fig. 3 of Paper I) have been binned to the spectral intervals of the radiative transfer scheme.
\begin{figure*}
  \centering
  \includegraphics[scale=0.60]{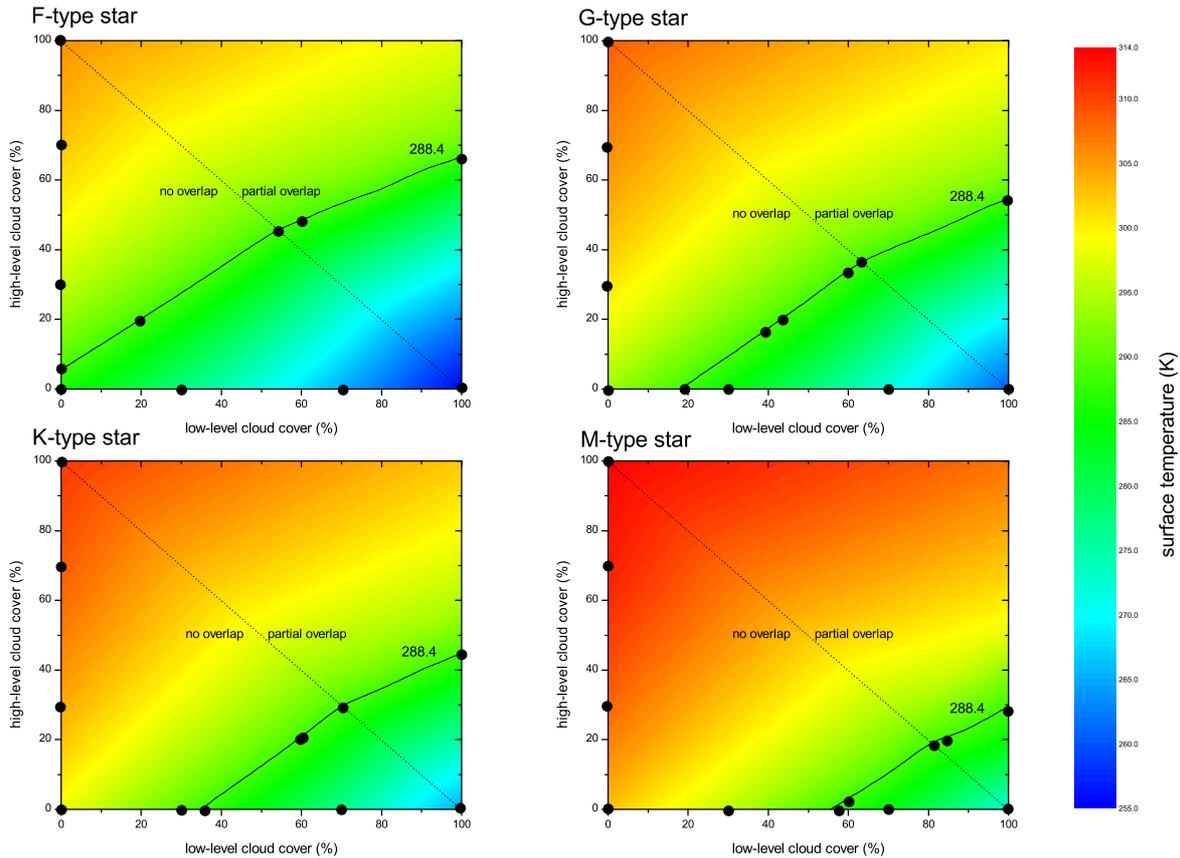}
  \caption{Surface temperatures of planets orbiting four different types of central stars as a function of the coverages of low and high-level clouds (see Paper I for a detailed discussion of these results). The solid lines on the contour plots represent parameters for which a mean Earth surface temperature of $288.4 \ \mathrm{K}$ is obtained. The dashed lines in the x-y-planes separate the regions between partial cloud overlap and no overlap. Black dots indicate the cloud coverages for which spectra are presented in this study at the relevant surface temperatures.}
  \label{surface}
\end{figure*}
We performed several calculations spanning the whole two-dimensional parameter space of possible cloud cover combinations for each stellar type in order to investigate the climatic effects of clouds in Earth-like planetary atmospheres (cf. Paper I). The measured value of the global mean surface albedo of the Earth (0.13) is used in our model. In contrast, the surface albedo had to be adjusted to mimic the effects of clouds in previous clear-sky model calculations \citep{Segura03,Segura05}. Such tuning is only possible, if the target surface temperature, for example the measured Earth global mean surface temperature (288.4 K), is prescribed. However in case of a model including clouds, as presented here, the surface temperature is a result of the calculations.

The contour plots shown in Fig. \ref{surface} summarise the resulting surface temperatures for the four central star types considered. A detailed description and discussion of these results is given in Paper I. The dots in Fig. \ref{surface} indicate the atmospheric model scenarios, for which the resulting low-resolution IR emission spectra are studied in this work. Special emphasis is put on the multi-layered cloud configurations which result in mean Earth surface conditions, namely a surface temperature of $288.4 \ \mathrm{K}$. The low-resolution spectra are directly obtained from the radiative transfer calculations of the atmosphere model.
Clouds, in addition to their effect on the IR emission spectra which are discussed in this article, also affect the reflected incident stellar radiation in the short wavelength part of the planetary spectrum. The investigation of the influence of clouds upon the reflection spectra of Earth-like planets will be done in future studies.

%\section{Earth reference model}
\section{Emission spectrum of Earth influenced by high and low-level clouds}
\label{sec_earth_ref}

For a qualitative comparison of our model calculations of (thermal) IR emission spectra of Earth-like planets influenced by low and high-level clouds with measurements of Earth, we study here first the resulting spectra obtained for the Earth model introduced in Paper I. This model reproduces the measurered global mean surface temperature ($288.4 \ \mathrm{K}$) with observed mean cloud coverages ($39.5\%$ low-level clouds, $15\%$ high-level clouds, and a $7\%$ overlap of both cloud layers). According to \citet{Warren07} the observed average Earth's total amount of cloud cover is $55\%$ ($68\%$) over land (ocean) yielding a global mean value of about $64\%$ which is much lower than the $48\%$ total cloud cover in the present model due to the omission of mid-level clouds. The clear sky calculation results in a surface temperature of $293 \ \mathrm{K}$, which is clearly too high (see Paper I for a detailed discussion of the climatic effects of clouds). 

Mid-level clouds (observed global mean cloud cover ca. $20\%$, \citet{Warren07}) had been omitted in the Earth model of Paper I, as most of them have been reported to be radiatively neutral, i.e. their greenhouse and albedo effect balance each other (see \citet{Poetzsch95} and Paper I for details). However, the neglect of the mid-level clouds yields less back-scattered shortwave and more outgoing longwave radiation at the top of the atmosphere (TOA) compared to the global energy budget of Earth\footnote{The corresponding radiation flux profiles for this Earth model are shown in Fig. 5 of Paper I.}. A comparison with measurements of \citet{Trenberth2009} shows a deviation about $13 \ \mathrm W \mathrm m^{-2}$ in the shortwave and longwave fluxes at the top of the atmosphere (see Table \ref{tab1}).
Presenting the IR emission spectra corresponding to the model described in Paper I only low and high-level clouds are considered in the present study. They represent the extreme cases of the effects of clouds on the spectrum of Earth. Oberservations of e.g. \citet{Tinetti2006a,Tinetti2006b} and \citet{Hearty2009} showed that the effects of mid-level clouds on the spectra of Earth are between these two extremes. Therefore, the important range of the cloud effects on the emission spectra are covered in our study of extrasolar planetary atmospheres. 

\begin{table}[t]
  \caption[]{Calculated radiation fluxes in comparison to measured values.}
  \label{tab1}
  \centering
  \begin{tabular}{l c c}
    \hline
    \noalign{\smallskip}
    Flux component & Measured Earth values & This work\\
    \hline
    \noalign{\smallskip}
    \textbf{Shortwave radiation} & & \\
    TOA down & $341.3 \ \mathrm W \mathrm m^{-2}$ & $340 \ \mathrm W \mathrm m^{-2}$ \\
    TOA up & $101.9 \ \mathrm W \mathrm m^{-2}$ & $89 \ \mathrm W \mathrm m^{-2}$ \\
    Planetary albedo & 0.3 & 0.26 \\
    \textbf{Longwave radiation} & & \\
    TOA up & $238.5 \ \mathrm W \mathrm m^{-2}$ & $251 \ \mathrm W \mathrm m^{-2}$ \\
    Surface down & $333 \ \mathrm W \mathrm m^{-2}$ & $350 \ \mathrm W \mathrm m^{-2}$ \\
    Surface up & $396 \ \mathrm W \mathrm m^{-2}$ & $392 \ \mathrm W \mathrm m^{-2}$ \\
    \hline
    \noalign{\smallskip}
  \end{tabular}
\end{table}

\begin{figure}
  \centering
  \resizebox{\hsize}{!}{\includegraphics{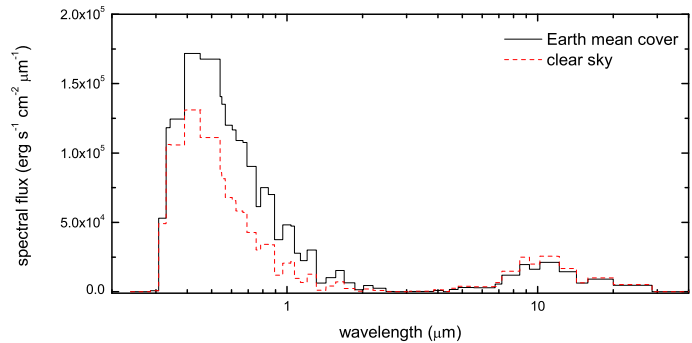}}
  \caption{Planetary spectra at the top of the atmosphere of the Earth model from near UV to IR wavelengths. The dashed line marks the clear sky case, the solid line indicates the results for a mean Earth cloud cover ($39.5\%$ low-level clouds, $15\%$ high-level clouds, and $7\%$ overlap of both cloud layers).}
  \label{earth_ref_spectrum}
\end{figure}

\begin{figure}
  \centering
  \resizebox{\hsize}{!}{\includegraphics{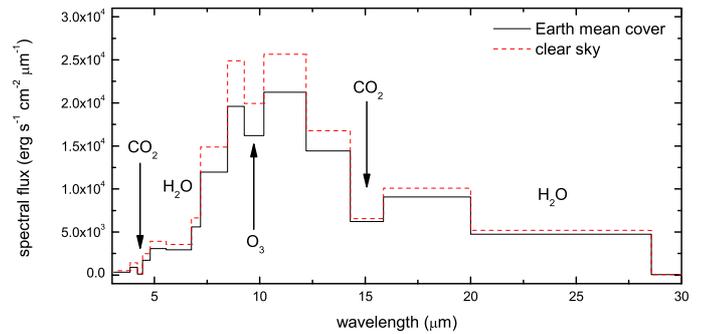}}
  \caption{Thermal emission spectra of the Earth model. The dashed line represents the clear sky case, the solid line marks the results for a mean Earth cloud cover ($39.5\%$ low-level clouds, $15\%$ high-level clouds, and $7\%$ overlap of both cloud layers). Chemical species which can be identified by absorption features are labelled (see text for details).}
  \label{earth_ref}
\end{figure}

Fig. \ref{earth_ref_spectrum} shows the related spectra covering the spectral wavelengths range from the near UV to the IR. In the visible range ($< 4 \ \mathrm{\mu m}$) we see a reflection spectrum composed of back scattered (and partly absorbed) incident solar radiation. In the IR wavelength range, we see an emission spectrum caused by thermal emission from the planetary surface and the atmosphere, influenced by atmospheric absorption and also scattering in the presence of clouds. 
The IR thermal emission spectra are depicted in greater detail in Fig. \ref{earth_ref}, revealing the most important spectral molecular absorption features which can be observered in low-resolution spectra.
The number of chemical species which can be identified by their respective absorption features in low-resolution spectra is limited for an Earth-like planet. In the wavelength region considered, $\mathrm{CO_2}$, $\mathrm{H_2 O}$, and $\mathrm{O_3}$ can be found. Clearly visible in the IR emission spectrum of the Earth are the $9.6 \ \mathrm{\mu m}$ absorption band of $\mathrm{O_3}$ and the $15 \ \mathrm{\mu m}$ band of $\mathrm{CO_2}$. Gaseous $\mathrm{H_2 O}$ has several rather broad (continuum) absorption bands due to rotation ($> 15 \ \mathrm{\mu m}$) and vibration ($5-8 \ \mathrm{\mu m}$).
According to \citet{Kaltenegger2009}, the spectral resolution, which is needed to identify the bands is about 20 for the $\mathrm{O_3}$ band. Absorption of $\mathrm{H_2 O}$ and $\mathrm{CO_2}$ (at $15 \ \mathrm{\mu m}$) can be seen at a much coarser resolution. Only the narrow absorption band of $\mathrm{CO_2}$ at $4.3 \ \mathrm{\mu m}$ requires a higher resolution of about 40.

Our results agree with the statement of \citet{Selsis2004} that only these three molecules would be effectively detectable on an extrasolar Earth with a realistic sensitivity and resolution. Obviously, the detection of other species requires a much higher spectral resolution.
\citet{Hearty2009}, for instance, additionally identified the molecules $\mathrm{N_2 O}$ and $\mathrm{CH_4}$ in their mid-IR AIRS observations of Earth. According to their findings the detection of stratospheric emission lines requires spectra with a resolution of $R=100$. However, the number of observable species can be different in case of non-Earth-like atmospheric compositions. 

Comparing the cloud-free clear sky and the mean Earth cloud cover spectra shown in Fig. \ref{earth_ref} reveals the principle impact of clouds on such emission spectra. The two main effects associated with the presence of clouds are the overall decrease of the outgoing IR flux and a dampening of spectral absorption features. The $\mathrm{O_3}$ band is strongly affected while other molecular bands, such as $\mathrm{CO_2}$, show only minor changes.
The decrease of the IR flux is partly a consequence of the lower surface temperatures as a result of the albedo effect of the low-level clouds on the one hand and the trapping of IR radiation in the lower atmosphere by the high-level clouds on the other. These findings are in very good agreement with measurements and calculations published by \citet{Tinetti2006a,Tinetti2006b} and \citet{Hearty2009}. Their studies investigated computed and measured high-resolution spectra of the Earth for different cloud conditions. Their resulting spectra show the same effects (decrease of the IR flux and reduced absorption features of especially $\mathrm{O_3}$ in the cases including clouds) as we found in our calculations. This suggests that our simplified radiative transfer model is therefore suitable for studying the basic effects of clouds on low-resolution spectra of Earth-like planets.

\section{Emission spectra  of Earth-like planetary atmospheres}
\label{sec_spectra}
In order to illustrate the basic effects of the two different cloud types on the planetary emission spectra, calculations using only single cloud layers are presented in the next subsection. The corresponding effects of multi-layered clouds for planets with mean Earth surface temperatures of $288 \ \mathrm{K}$ are discussed subsequently.
\begin{figure*}
  \centering
  \resizebox{\hsize}{!}{\includegraphics{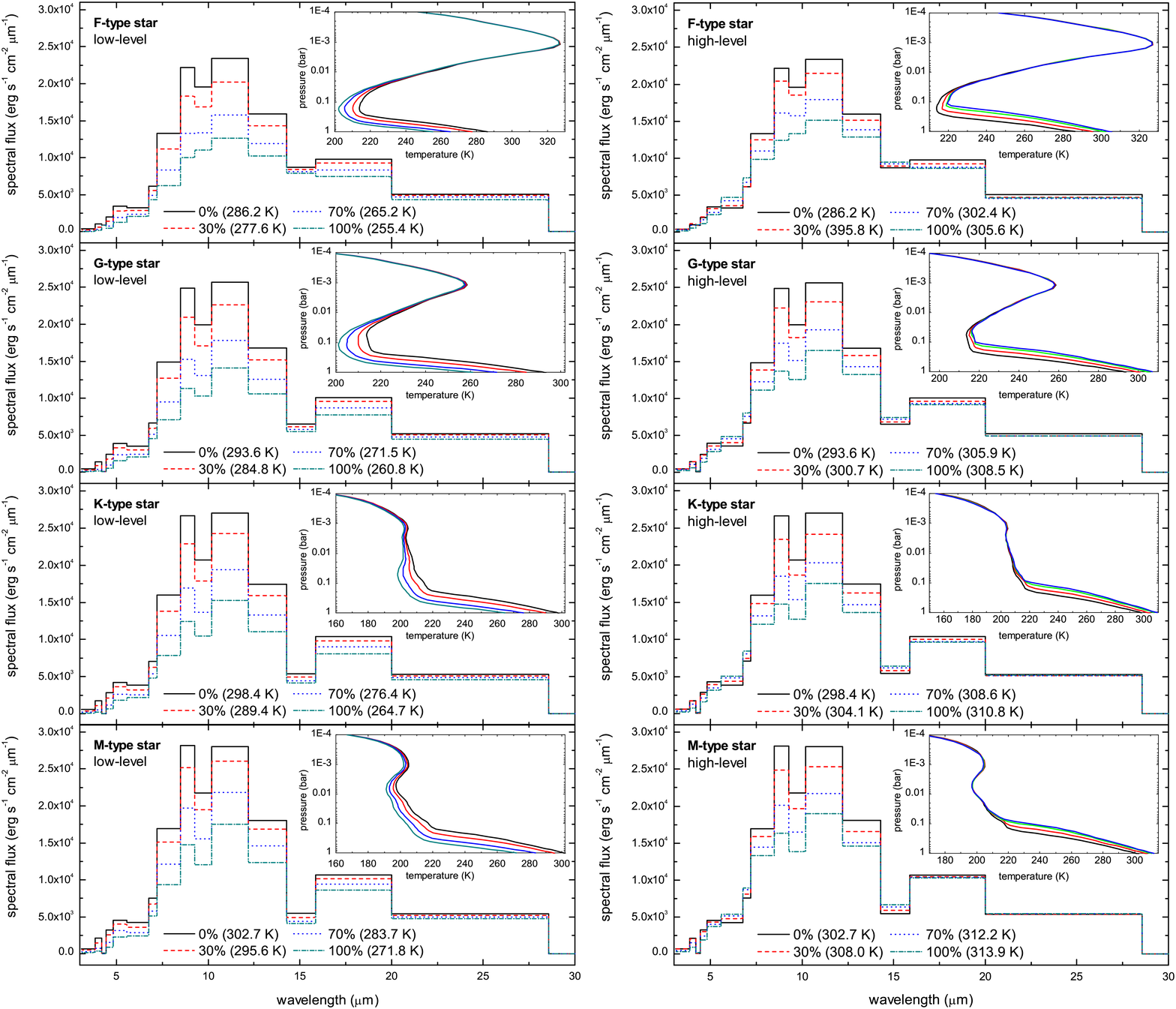}}
  \caption{Thermal emission spectra influenced by single low-level clouds (\textit{left diagram}) and high-level clouds (\textit{right diagram}). The resulting spectra are shown for each stellar type and four different cloud covers (solid lines: $0\%$, dashed lines: $30\%$, dotted lines: $70\%$, and dashed-dotted lines: $100\%$ coverage of the respective cloud type). The planetary surface temperatures resulting from these cloud coverages are given in parenthesis. The corresponding temperature profiles are shown in the inset plots.}
  \label{spectra_single_layer}
\end{figure*}
\subsection{Single high-level ice and low-level water droplet clouds}
\label{sub_basic_effects}
\begin{figure*}
  \centering
  \resizebox{\hsize}{!}{\includegraphics{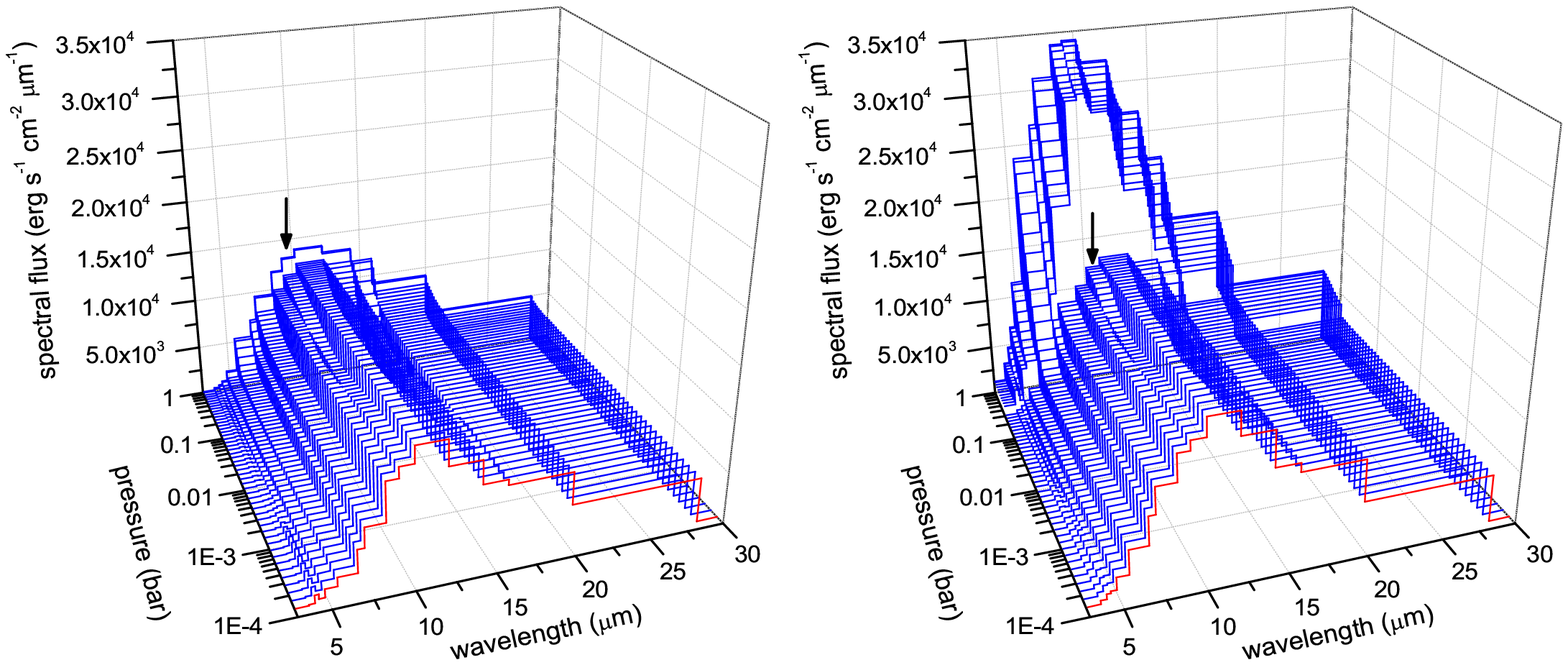}}
  \caption{Spectral tomography of the thermal emission spectra from the surface to the top of the atmosphere of an Earth-like planet orbiting the F-type star. \textit{Left diagram}: $100\%$ single low-level cloud cover ($T_{\mathrm{surf}} = 255.4 \ \mathrm{K}$), \textit{right diagram}: $100\%$ coverage of single high-level clouds ($T_{\mathrm{surf}} = 305.6 \ \mathrm{K}$). The position of the ozone band is marked by arrows.}
  \label{f3d}
\end{figure*}
Low-resolution spectra of Earth-like planetary atmospheres for single high-level ice and low-level water clouds have been obtained for each stellar type. The results are shown in Fig. \ref{spectra_single_layer} for three different single cloud coverages ($30\%$, $70\%$, and $100\%$) and the respective clear sky cases. The inset plots show the respective temperature profiles.

Even if no clouds are present in the atmospheres and the same atmospheric chemical composition is assumed the resulting emission spectra differ for each stellar type due to the direct influence of the central star. For the F-type star case, the overall outgoing planetary IR flux is the lowest, while the planet around the M-type star shows the highest IR emission. This is directly connected to the corresponding surface temperatures of the planets, which are the lowest in the case of the F-type star and almost $20 \ \mathrm{K}$ higher for the M-type star planet (cf. Fig. \ref{surface} and Paper I). 

However not only the overall IR flux is different, also the absorption features of $\mathrm{O_3}$, $\mathrm{CO_2}$, and $\mathrm{H_2O}$ change due to the central star. The depths of the absorption bands of e.g. the planet around the F-type star are rather small compared to the M star case. In particular, the $\mathrm{CO_2}$ absorption band at $4.3 \ \mathrm{\mu m}$ cannot be seen for the F-type star case in the low-resolution spectrum (cf. Fig. \ref{spectra_single_layer}). The different clear sky spectra are a direct consequence of the different atmospheric temperature profiles. The planet around the F-type star has a very large temperature inversion in the upper atmosphere \citep[see also][]{Segura03,Grenfell07}. This temperature inversion leads to enhanced emission in the $\mathrm{O_3}$ and $\mathrm{CO_2}$ bands, thereby effectively reducing the depths of their absorption features. The planets around the M-type and K-type stars on the other hand show no large atmospheric temperature inversion which leaves the depths of the absorption features almost unaltered from emission (see \citet{Segura03,Segura05} for a description of these temperature profile characteristics. Note however, that these cloud free spectra are not directly comparable to those of \citet{Segura03,Segura05} because of differences in the atmospheric modelling (e.g. different treatments of the atmospheric chemistry and the incident stellar spectra) as outlined above.

The resulting planetary thermal emission spectra for different stellar types are profoundly affected by the cloud layers. In general, the emitted overall IR radiation flux decreases and the molecular absorption bands ($\mathrm{O_3}$, $\mathrm{CO_2}$, and $\mathrm{H_2O}$) are dampened with increasing cloud cover. These fundamental changes in the spectra occur for both cloud types as shown in Fig. \ref{spectra_single_layer}. The most extreme effect occurs again in the case of the F-type star planet, where the absorption features of $\mathrm{O_3}$ and $\mathrm{CO_2}$ cannot be distinguished anywhere in the low-resolution spectra for high (single) cloud coverages (see Fig. \ref{spectra_single_layer}). The F-star planet spectra appear more similar to emission spectra than absorption spectra, even though the amount of $\mathrm{CO_2}$ and $\mathrm{O_3}$ remains unchanged in the calculations. In the spectra of planets orbiting the other types of central stars, the features of $\mathrm{CO_2}$ and $\mathrm{O_3}$ are also strongly affected, but remain visible even at full cloud cover. The major absorption band of $\mathrm{CO_2}$ around $15 \ \mathrm{\mu m}$ itself is not directly influenced very much by the cloud layers, as it is optically very thick throughout the whole troposphere. Since however the adjacent bands are strongly affected, its detectability would be more difficult, especially for the F-type star.

Figure \ref{spectra_single_layer} implies that both types of clouds are associated with a decrease in the emerging IR flux with increasing cloud cover. However, the physical reasons for this effect are totally different for each cloud type. As already discussed in Paper I, low-level clouds exhibit a net albedo effect, which results in lower surface temperatures with increasing cloud coverage. This in turn leads to less emitted IR radiation at the surface and thus to a decrease of the outgoing IR flux at the top of the planetary atmosphere. By contrast high-level clouds show a net greenhouse effect resulting in higher surface temperatures. However, since this greenhouse effect also traps the IR radiation in the lower atmosphere, the finally emitted IR spectrum at the top of the atmosphere is also reduced.

Intuitively one would expect that clouds have no direct effect on the $\mathrm{O_3}$ absorption feature because the cloud layers are located in the troposphere, while the $\mathrm{O_3}$ band originates in the upper atmosphere. Our results however indicate that water and ice clouds have a strong impact on this absorption feature. For each type of central star it can clearly be seen (Fig. \ref{spectra_single_layer}), that the spectral bands adjacent to the $\mathrm{O_3}$ band (especially the adjacent band at smaller wavelengths) are more affected by clouds than the $\mathrm{O_3}$ band itself. This finally leads to a reduced contrast of the $\mathrm{O_3}$ absorption band
in low-resolution spectra, even though the ozone absorption feature itself is in principle still present. The optical properties of our cloud model (cf. Fig. 1 of Paper I) show that the $\mathrm{O_3}$ absorption band is located in a minimum of the extinction optical depths of both cloud types around $\sim 10 \ \mathrm{\mu m}$, while the adjacent band at smaller wavelengths has a higher optical depth. This results in a stronger decrease of the radiation flux at this wavelength band compared to the adjacent $\mathrm{O_3}$ band when passing through the cloud layers. Consequently, the widely used argument that clouds cannot affect spectral features originating from parts of the atmosphere \textit{above} the cloud layer(s) is not valid for their detectability at low resolution. 

Fig. \ref{f3d} illustrates the spectral tomography of the thermal emission spectra from the planetary surface to the top of the atmosphere of an Earth-like planet around the F-type star. The figure compares the two single cloud cases with maximum coverage. 

The disappearance of the $\mathrm{O_3}$ feature (at $9.6 \ \mathrm{\mu m}$) throughout the atmosphere can be seen in Fig. \ref{f3d} for both cases. \textit{Inside the atmosphere} the feature is still visible at pressures of about $0.1 \ \mathrm{bar}$. However, the absorption band is subsequently filled (i.e. becoming weaker) by emission due to the atmospheric temperature inversion\footnote{This mechanism is limited by the renewed temperature decrease in the upper part of the planetary atmosphere (see Fig. \ref{f3d}).}. This leads ultimately, in combination with the previously discussed optical depth effect (clouds affect the adjacent band at smaller wavelengths more than the $\mathrm{O_3}$ band itself) to the disappearance of the feature at the top of the atmosphere of the F-star planet in both low-resolution spectra.

As can be clearly noticed, the low-level cloud (at $\sim 0.8 \ \mathrm{bar}$) has almost no direct effect on the outgoing IR radiation (left diagram of Fig. \ref{f3d}). The low-level clouds absorb a part of the outgoing IR flux, but they also emit thermal IR radiation depending on their own distinct temperature at the same time. Since these clouds are located near the surface, their emission temperature is comparable to the surface temperature of the planet. Therefore, no large effect on the outgoing IR radiation is apparent. 
However, due to the high optical depth of the low-level cloud layer, the planetary surface is effectively concealed, including possible characteristic surface emission features. Since the thermal emission of the planetary surface is described by black-body radiation of the corresponding surface temperature in our model, these kinds of effects cannot be investigated in this study \citep[see e.g.][for discussion of these effects for Earth]{Tinetti2006b}. Even though low-level clouds exhibit no large direct impact on the IR emission spectra, they have a huge influence on the surface temperature (albedo effect) and therefore affect the emitted IR spectra indirectly.
\begin{figure*}
  \centering
  \resizebox{\hsize}{!}{\includegraphics{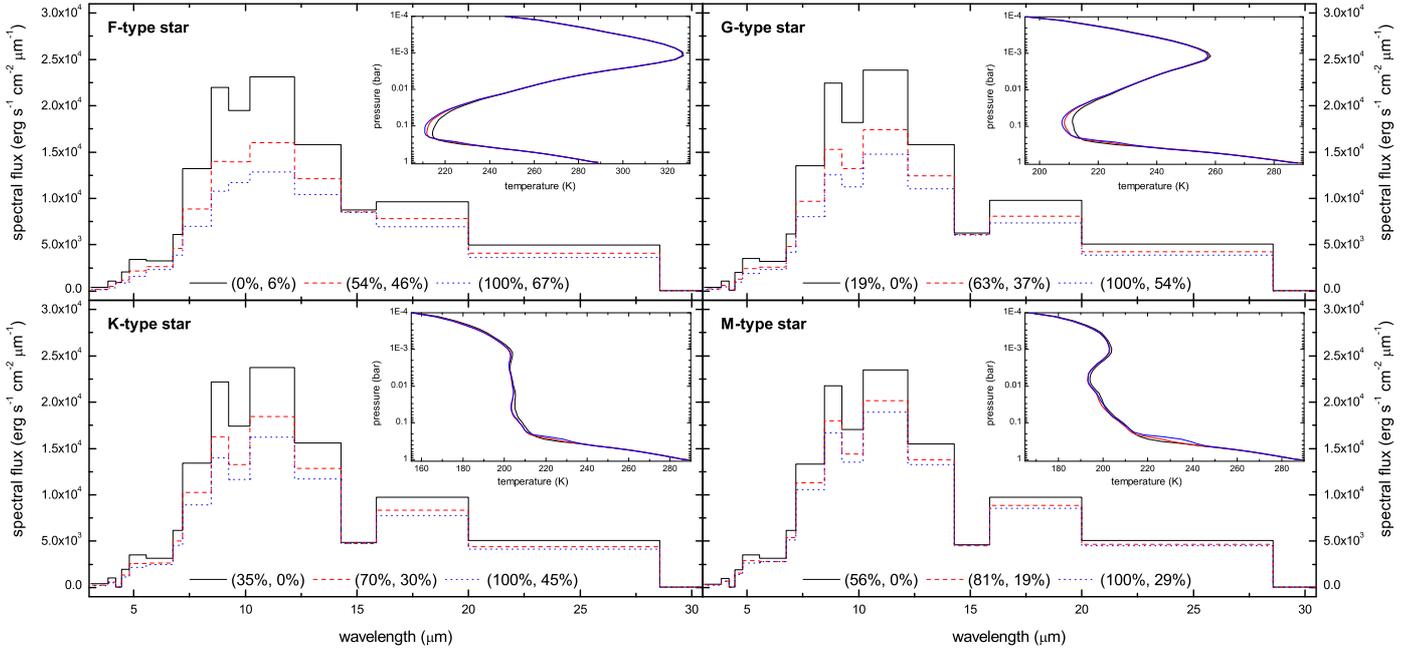}}
  \caption{Planetary thermal emission spectra influenced by multi-layered clouds with three different cloud cover combinations yielding surface temperatures of $288.4 \ \mathrm{K}$ in all cases. The inset plots show the corresponding temperature profiles. The cloud coverages used are given in parenthesis (low-level cloud cover, high-level cloud cover).}
  \label{288_ir}
\end{figure*}
The high-level ice clouds on the other hand have a very large direct effect on the thermal emission spectrum (right diagram of Fig. \ref{f3d}). Just like the low-level clouds they affect the outgoing IR radiation by absorption, scattering, and emission. The emission temperatures of these high-level clouds are in general much lower than the surface temperature which leads to the trapping of IR radiation in the lower atmosphere (the greenhouse effect already discussed) and thus to a decrease in the outgoing IR radiation. The huge drop of the radiation flux at the location of the ice cloud layer (at $\sim 0.3 \ \mathrm{bar}$) in the atmosphere can clearly be seen in Fig. \ref{f3d} (right diagram). However, despite the very different physical processes the emission spectra for the two cloud types are quite similar in shape at the top of the atmosphere (Fig. \ref{f3d}).

\subsection{Effects of multi-layered clouds}

In this subsection we extend our study to the combined effects of multi-layered clouds. In particular we focus on the subset of atmospheric models with multi-layered clouds from Paper I, which are related to mean Earth surface conditions (i.e. surface temperatures of $288.4 \ \mathrm{K}$). As already discussed, various combinations of cloud covers of high and low-level clouds can result in mean Earth surface temperatures. The possible range of combinations however depends on the spectral properties of the host star. Fig. \ref{surface} summarises our results with respect to the surface temperatures due to different cloud coverages obtained in Paper I. The atmospheric models with $\mathbf{T_\mathrm{surf}=}288.4 \ \mathrm{K}$ are marked by the solid iso-lines. Even though the surface temperatures are the same in all of these cases the different amount of cloud coverages of the two cloud types and resulting differences in the atmospheric temperatures profiles lead in turn to changes in the emission spectra.

\subsubsection{Characteristic molecular features}

For each {planet-star scenario}, three different cloud cover combinations have been chosen. The corresponding thermal IR spectra are shown in Fig. \ref{288_ir}. The inset plots of Fig. \ref{288_ir} show the atmospheric temperature profiles. The temperature profiles correspond qualitatively to \citet{Segura03,Segura05}. There are however some quantitative differences because of different treatments of the atmospheric chemistry, clouds, radiative transfer, and the distances of the planets to their host stars (see Paper I for a detailed discussion).

Fig. \ref{288_ir} suggests that the differences between the temperature profiles for a planet/star configuration are rather small despite the large variations of the cloud coverages. The maximum deviations are about $5 \ \mathrm{K}$ in each case. The small changes in the gas opacities correlated with these temperature variations can not explain the huge changes in the related emission spectra. Therefore, the variations in the emission spectra of Fig. \ref{288_ir} are likely caused by the effects of clouds.

Depending on the prescribed multi-layered cloud coverage, the resulting spectra show large differences even though in each calculation the same mean Earth surface temperature is obtained (Fig. \ref{288_ir}). Despite the fact, that the atmospheric profile of e.g. $\mathrm{O_3}$ is the same for all models, the possibility to detect its spectral feature in a low-resolution spectrum is severely affected by absorption and scattering of the outgoing IR radiation by cloud particles. The $\mathrm{O_3}$ feature is not apparent in the spectrum for higher cloud coverages in the case of the F-type star (see Sect. \ref{sub_basic_effects}), while for the other kinds of central stars this feature is still present, though strongly dampened. Absorption bands of other important chemical species, such as $\mathrm{H_2O}$ or $\mathrm{CO_2}$ are also affected, either directly or indirectly. As already discussed in Sect. \ref{sub_basic_effects} the direct influence of the cloud layers on the  $15 \ \mathrm{\mu m}$ absorption band of $\mathrm{CO_2}$ is not very effective. However, since the adjoining bands are strongly affected by clouds it would be difficult to observe this feature in low-resolution spectra, especially again in case of the F-type star.

\begin{figure}
  \centering
  %\resizebox{\hsize}{!}{\includegraphics{plots/all_ll_ir.eps}}\\
  %\resizebox{\hsize}{!}{\includegraphics{plots/all_hl_ir.eps}}
  \resizebox{\hsize}{!}{\includegraphics{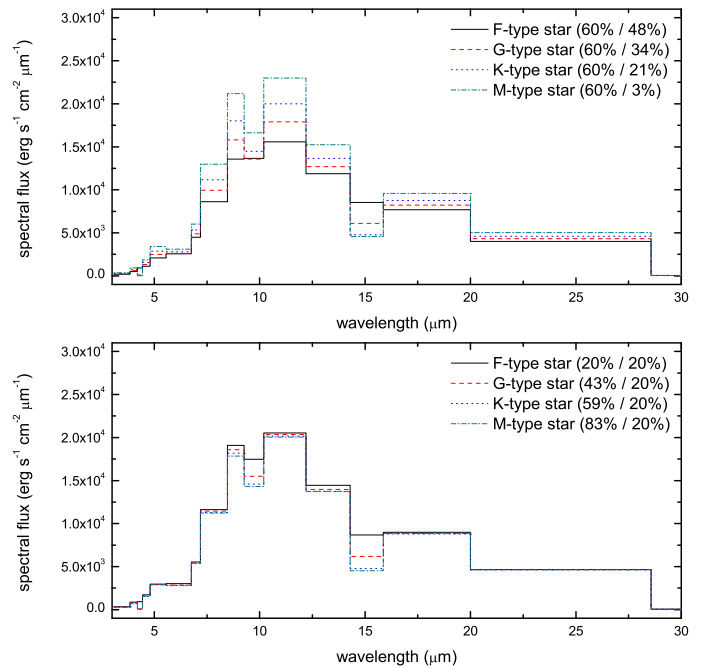}}
  \caption{Multi-layered cloud cover scenarios for  $288.4 \ \mathrm{K}$ surface temperature (low-level cloud cover / high-level cloud cover). \textit{Upper diagram}: fixed low-level cloud cover of $60\%$ and adjusted high-level cloud coverage (see Fig. \ref{surface}), \textit{lower diagram}: fixed high-level cloud cover of $20\%$ and adjusted low-level cloud cover (see Fig. \ref{surface}).}
  \label{all_288}
\end{figure}

The emission spectra of all four central star scenarios are compared in Fig. \ref{all_288} for two different cloud covers. In the first case, a low-level cloud of $60\%$ is assumed for each type of planet and the corresponding high-level cloud cover is adjusted to achieve mean Earth surface conditions. For the second case, a uniform high-level cloud coverage of $20\%$ is assumed and the corresponding amount of low-level clouds is adjusted. 
In the first scenario, the spectra are obviously affected due to the required high-level cloud cover. The more high-level clouds are present the less IR flux is emitted at the top of the atmosphere and the more the spectral features are dampened. For the second case, the shape of the emission spectra look almost equal despite the very different amount of low-level clouds. Noticeable changes can be found in the $9.6 \ \mathrm{\mu m}$ $\mathrm{O_3}$ and the $15 \ \mathrm{\mu m}$ $\mathrm{CO_2}$ bands. These differences however are not the results of the different cloud covers, but of the different temperature profiles as already discussed in Sect. \ref{sub_basic_effects}. 

From the direct comparison of the two scenarios shown in Fig. \ref{all_288} it is obvious that the high-level ice clouds have the strongest impact on the thermal emission spectra for planets with the same surface temperature. This can be understood by the fact that the low-level clouds emit thermal radiation at almost the same temperature as the surface (previously discussed in Sect. \ref{sub_basic_effects}). This limits their influence on the spectra in these particular cases because the surface temperature is the same for all considered scenarios. The high-level clouds on the other hand are located higher up in the atmosphere and emit at temperatures generally much colder than the surface. Their influence on the spectra therefore directly depends on their respective coverage (see also Sect. \ref{sub_basic_effects}).

\subsubsection{IR continuum radiation and related planetary temperatures}

Thermal emission spectra can in principle be used to obtain information about atmospheric or surface temperatures of planets. Usually black-body radiation curves are fitted to the spectral shape of observed thermal emission spectra. Clouds conceal the surface below them as a function of their optical depth, but also act as emitters of thermal radiation. 
Since clouds emit at their own radiative equilibrium temperature, the temperatures derived by fitting of black-body curves are in general not directly connected to the planetary surface temperature if high-level clouds are present in the atmosphere.

\begin{figure}
  \centering
  \resizebox{\hsize}{!}{\includegraphics{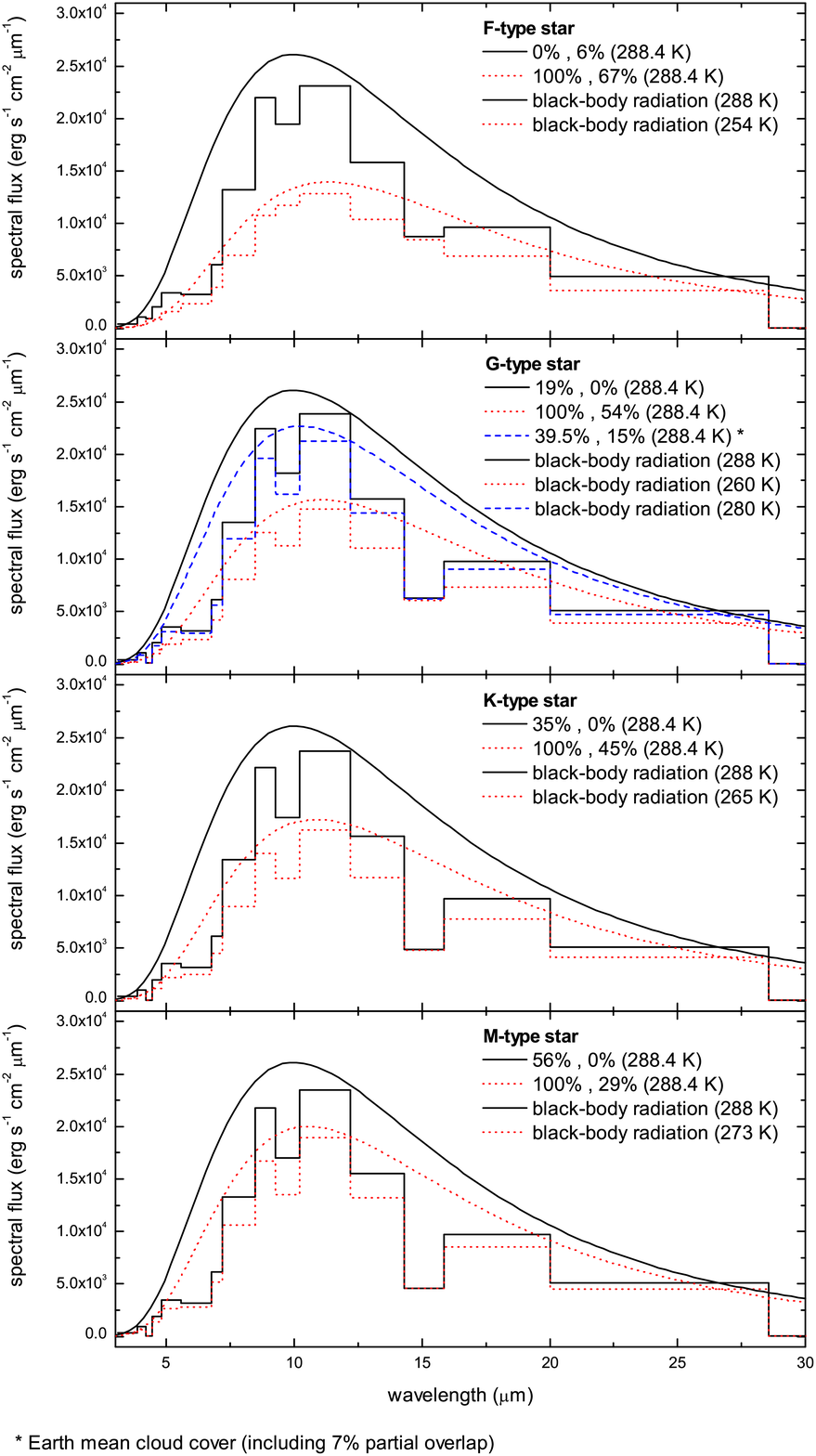}}
  \caption{Derivation of planetary temperatures from the shapes of IR emission spectra. The black-body radiation curves have been obtained from a best fit by eye on selected spectra of Fig. \ref{288_ir} with the exception of the $288.4 \ \mathrm{K}$ curves shown for comparison. The cloud coverages used are indicated: low-level cover, high-level cloud cover. The corresponding temperatures (surface temperatures and derived temperatures of the black-body radiation) are given in parenthesis.}
  \label{blackbody_all}
\end{figure}
Fig. \ref{blackbody_all} shows two spectra, namely for a small and a large cloud cover, for each of the central star scenarios.
In addition the results of the Earth model (Sect. \ref{sec_earth_ref}) are shown. In all cases the surface temperature equals $288.4 \ \mathrm{K}$. The temperatures derived from the fitting of black-body radiation curves, however, can be very different. For cases with a small (or even no) coverage of high-level clouds, the black-body radiation curve with $288 \ \mathrm{K}$ fits quite well. Even a large amount of low-level clouds, as in the case of the M-type star ($56\%$ low-level cloud cover), has no large effect upon the emission spectrum. The reason for this behaviour is of course the already discussed effect of the low-level clouds which emit black-body radiation IR spectra comparable to the surface temperature. With increasing amount of high-level clouds, the temperatures which can be inferred by examining the spectra generally decrease, since these clouds are at the upper boundary of the troposphere and emit at temperatures lower than the surface temperature. 

For all considered scenarios in Fig. \ref{blackbody_all} where a large amount of high-level clouds is present the temperatures obtained from the spectra would suggest inhabitability if these temperatures would be interpreted as the respective surface temperatures. Since the types and corresponding amount of clouds and therefore their influence on the spectra is not known a priori the information about the surface temperatures derived from low-resolution emission spectra can be misleading, especially for large amounts of high-level clouds. However, the derived black-body temperatures would equal the temperature at the tropopause for $100\%$ high-level cloud cover, which agrees with the findings of \citet{Tinetti2006a} for Earth. Clouds are therefore clearly limiting the ability of obtaining temperature information about the planetary surface from an observed low-resolution spectrum.

\section{Summary}

Our previously developed parametrised cloud description (Paper I) coupled with a one-dimensional radiative-convective climate model was used here to study the impact of multi-layered clouds on the thermal emission spectra of Earth-like extrasolar planets.

We showed that both cloud types, single low-level water and single high-level ice clouds, ultimately lead to similar changes in the thermal emission spectra. As the cloud cover of either type increases, the overall IR emissions decreases and spectral absorption features are dampened. However, the physical effects caused by the the different types of clouds are not the same. In the case of low-level water clouds the decrease of the surface temperatures (albedo effect, cf. Paper I) is responsible for the decrease of the IR emission, while in the case of the high-level ice clouds the trapping of IR radiation in the lower atmosphere (greenhouse effect) leads to the decrease in the outgoing IR flux.
 
The effects of multi-layered clouds have been studied on model scenarios where the cloud coverage leads to mean Earth surface temperatures for each central star. In these cases the low-level clouds have no large effect on the thermal emission spectra, because these clouds emit IR radiation at temperatures comparable to the planetary surface. Since however in all of the models considered the surface temperature was the same this effect was therefore weak. The high-level clouds on the other hand emit IR radiation at considerably lower temperatures which leads to a decrease in the outgoing IR radiation and dampening of spectral absorption features. This dampening is not limited to absorption features originating from below the cloud layers, but can also be found for the $9.6 \ \mathrm{\mu m}$  $\mathrm{O_3}$ band which forms mostly in the upper atmosphere. The strongest effects on the emission spectrum is found for the planet orbiting the F-type star where no characteristic molecular absorption features can be distinguished in the low-resolution spectrum for high cloud coverages. We note, however, that in most other cases, in particular for planets around M dwarfs, the interesting absorption bands remain present even at high cloud coverages.

Clouds also affect the ability to obtain surface temperatures from low-resolution spectra by fitting black-body radiation curves to the spectral shape of the emission spectra. With increasing amount of high-level clouds, the derived temperatures are much lower than the real surface temperature. Thus, clouds affect the ability to obtain information about the planetary surface temperature from low-resolution spectra. Planets with observed surface temperatures somewhat below mean Earth conditions should, therefore, not be discarded too quickly from the list of potentially habitable planets.
To sum up, clouds can have a significant impact on the measurement of apparent temperatures of planets and the detectability of the characteristic spectral signatures in the infrared used as tracers of habitability and life. 
%Consequently, our findings support the conclusion drawn by \citet{Hearty2009}, that the cloud amount will be the most observable characteristic of extrasolar Earth-like planets in the mid-IR.

\begin{acknowledgements}
  The authors would like to thank J.L. Grenfell for his suggestions improving the manuscript.
  This work has been partly supported by the research alliance \textit{Planetary Evolution and Life} of the Helmholtz-Gemeinschaft.
\end{acknowledgements}

\bibliographystyle{aa} % style aa.bst
\bibliography{references}

\end{document}